# Coherent topological polariton laser

T. H. Harder[1], M. Sun[2,3], O. A. Egorov[4], I. Vakulchyk[2,3], J. Beierlein[1], P. Gagel[1], M. Emmerling[1], C. Schneider[1], U. Peschel[4], I. G. Savenko[2,3], S. Klembt[1] and S. Höfling[1,5]

[1]Technische Physik, Wilhelm-Conrad-Röntgen-Research Center for Complex Material Systems, and Würzburg-Dresden Cluster of Excellence ct.qmat, Universität Würzburg, Am Hubland, D-97074 Würzburg, Germany
[2]Center for Theoretical Physics of Complex Systems, Institute for Basic Science (IBS), Daejeon 34126, Korea
[3]Basic Science Program, Korea University of Science and Technology (UST), Daejeon 34113, Korea
[4]Institute of Condensed Matter Theory and Optics Friedrich-Schiller-University Jena, Max-Wien-Platz 1, D-07743 Jena, Germany
[5]SUPA, School of Physics and Astronomy, University of St. Andrews, KY 16 9SS, United Kingdom

**Topological concepts have been applied to a wide range of fields in order to successfully describe the emergence of robust edge modes that are unaffected by scattering or disorder. In photonics, indications of lasing from topologically protected modes with improved overall laser characteristics were observed. Here, we study exciton-polariton microcavity traps that are arranged in a one-dimensional Su-Schrieffer-Heeger lattice and form a topological defect mode from which we unequivocally observe highly coherent polariton lasing. Additionally, we confirm the excitonic contribution to the polariton lasing by applying an external magnetic field. These systematic experimental findings of robust lasing and high temporal coherence are meticulously reproduced by a combination of a generalized Gross-Pitaevskii model and a Lindblad master equation model. Thus, by using the comparatively simple SSH geometry, we are able to describe and control the exciton-polariton topological lasing, allowing for a deeper understanding of topological effects on microlasers.**

**Introduction**

Since the discovery and systematic description by Kosterlitz, Thouless [1] and Haldane [2], topological phenomena have covered a wide variety of physical systems with the successive emergence of the concept of topological insulators. The latter are characterized by propagating edge states of reduced dimensionality that occur at the boundary between areas with different topological invariants. An outstanding property of these edge states is their robustness against scattering by defects ensured by topological protection. While topological effects have been first discovered experimentally in electronic systems in the quantum Hall effect regime [3], they have, among others, been proposed [4-8] and subsequently realized in the fields of photonics [9] and polaritonics [10,11]. Here, topologically protected laser modes have attracted considerable interest [12-17]. In this work, we present a well-controlled platform based on a system of coupled traps in a semiconductor microcavity hosting exciton-polaritons (later *polaritons*) [18,19]. Topological phenomena in polaritonic systems are drastically different from those in other platforms due to the hybrid light-matter nature of these quasi-particles, which allows them to undergo a condensation-like transition at elevated temperatures that leads to the emission of a coherent light [20].

Exciton-polaritons arise from the strong coupling between the photonic mode in a microcavity and an excitonic mode. Due to their part-light, part-matter composition, polaritons feature a unique set of properties including a small effective mass inherited from the photonic component, as well as the ability to interact with each other and be susceptible to external magnetic fields due to the excitonic component. Following the first demonstration of the strong coupling regime [18], this set of properties has led to the observation of a transition to a macroscopic occupation of a single polariton ground state under high pumping powers, referred to as *polariton condensation* [20].

Recent advances in technological control allow for the development of various trapping potentials for polaritons and polariton condensates. While the first realizations of polariton lattices based on metal layers on top of the cavity were still characterized by low confinement potentials [21], lattices based on coupled, etched micropillars overcame this limitation [22]. This control over polariton potentials has led to new implementations of non-trivial topology in one-dimensional [10] and two-dimensional lattices [11] that have attracted particular interest in the context of topological lasing [12-17] based on the coherent, laser-like emission of polariton condensates [23-25].

Due to its convincingly simple geometry, the Su-Schrieffer-Heeger (SSH) model [26,27], originally developed to describe the alternating bond pattern in polyacetylene, has evolved as one of the most significant tools for topology in numerous platforms, such as photonics [15,28] and polaritonics [10,29,30]. Here, we use this model to advance topological polaritonics and introduce a technological platform to support and control highly coherent lasing from polaritonic topological edge modes. Additionally, our results allow to accurately tune the spectral position of the topological mode in the gap.

## Results

**Experimental platform and implementation of a topological SSH model**

Our technological platform is based on the etch-and-overgrowth process (EnO). Here, the microcavity growth is interrupted after the bottom distributed Bragg reflector (DBR) and cavity with integrated quantum wells are finished. The trapping landscape is etched directly into the top of the cavity. Subsequently, the sample is cleaned with a hydrogen plasma and overgrown with a top DBR [31-33]. This fabrication process enables us to accurately control the confinement potential, that is directly linked to the etch depth, as well as the inter-site coupling in any given lattice geometry [34,35]. Another major advantage of this trapping technique is the prevention of exposed etched sidewalls [32] when compared to standard micropillar etching.

In order to create a topological edge defect in a one-dimensional chain, we implement an orbital SSH model [10,29]. The SSH model describes a dimerized chain with two sites per unit cell, parametrized by different intra- ($v$) and inter-cell ($w$) hopping coefficients. In the tight-binding limit, that is relevant for our system [35], the Hamiltonian can be written as

$$\hat{H} = v \sum_{m=1}^{N} (|m, B\rangle\langle m, A| + h.c.) + w \sum_{m=1}^{N-1} (|m+1, A\rangle\langle m, B| + h.c.),$$

where $N$ denotes the number of unit cells, $A$ and $B$ the sites in the unit cell, and $v$ and $w$ the respective hopping amplitudes [27]. This Hamiltonian reveals two topologically distinct phases for the cases $v < w$ and $v > w$. The topological difference can be understood by calculating the difference in phase winding

$$W = \frac{1}{2\pi} \int_{BZ} \frac{\partial \varphi(k)}{\partial k} dk,$$

where the winding $W$ of the phase $\varphi(k)$, corresponding to the geometrical phase term $e^{-i\varphi(k)}$ of the eigenfunctions of the $A$ and $B$ sublattices in momentum space, is calculated across the Brillouin Zone (BZ). A weakly bound edge pillar ($v < w$) leads to a topological defect and a bulk winding number of $W = 1$, whereas the opposite case remains topologically trivial. In the orbital SSH model this concept is applied to the $P$-mode of a pillar zigzag chain (see Figs. 1a, b) and treats the $P_x$ and $P_y$ sub-modes as two individual implementations of the SSH model. The orientation of these orbital modes naturally leads to the difference in coupling strengths $v \neq w$.

In Fig. 1c, the real space mode spectrum of a single, uncoupled polariton trap with a diameter of $d$ = 3.5μm is depicted. Adjacent to the spectrum, real space images of the $S$- and $P$-modes that were extracted from a mode tomography are presented. The characteristic patterns of these modes intuitively illustrate that a zigzag chain of S-modes leads to an entirely symmetric and topologically trivial case, whereas the distinct orientation of the $P_x$ and $P_y$ sub-modes results in a topological defect for the sub-mode that features a small mode overlap and, thus, a weak bond at the end of a zigzag chain. In Figure 1d, images of the S-mode as well as the topological defect of the $P_y$ sub-mode of a zigzag chain with trap diameters of d = 3.5µm and a reduced trap distance of $v = a/d = 0.8$, with $a$ denoting the center-to-center distance between adjacent traps, that were obtained under non-resonant laser excitation, are presented. While the trivial *S*-mode nicely reveals the sample structure, the *P* mode exhibits the expected topological defect edge mode. Our technological implementation of the SSH-Hamiltonian is based on a patterned array of buried polariton traps (see Fig. 1a) that are characterized by a confinement potential of 11.5meV and a Rabi splitting of 4.5meV (for further details on the sample, see Methods).

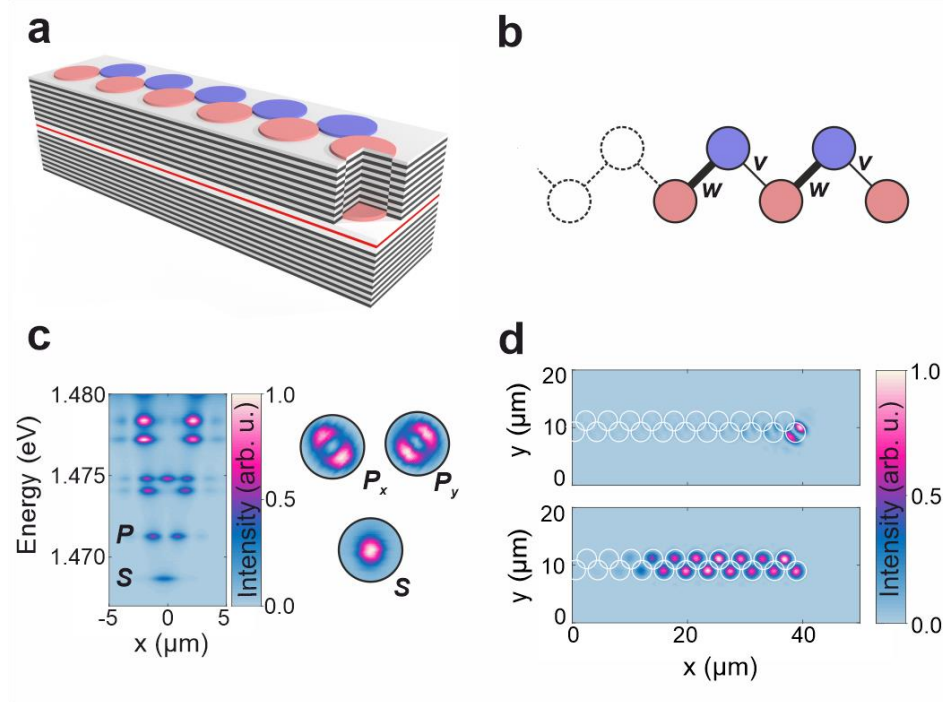

**Figure 1 | Schematic representations of device design and dimerization in the orbital SSH model, leading to a topological defect. a,** Schematic drawing of a polariton zigzag chain consisting of coupled microcavity traps created by etching of the cavity layer and subsequent overgrowth. **b,** Schematic representation of the chain dimerization with $v < w$ leading to a winding number of $\mathcal{W} = 1$ and a distinct topological edge mode localized to the last trap at the end of the chain. **c,** Spectrally resolved mode spectrum of an uncoupled polariton trap showing the rotationally symmetric S-mode and the P-mode consisting of the $P_x$ and $P_y$ sub-modes. **d,** Real space mode tomography showing the topological SSH edge mode (top) and the trivial S-mode (bottom) under non-resonant laser excitation with an elongated large spot on a zigzag chain with trap diameters of $d = 3.5\mu m$ and a reduced trap distance of $v = 0.8$.

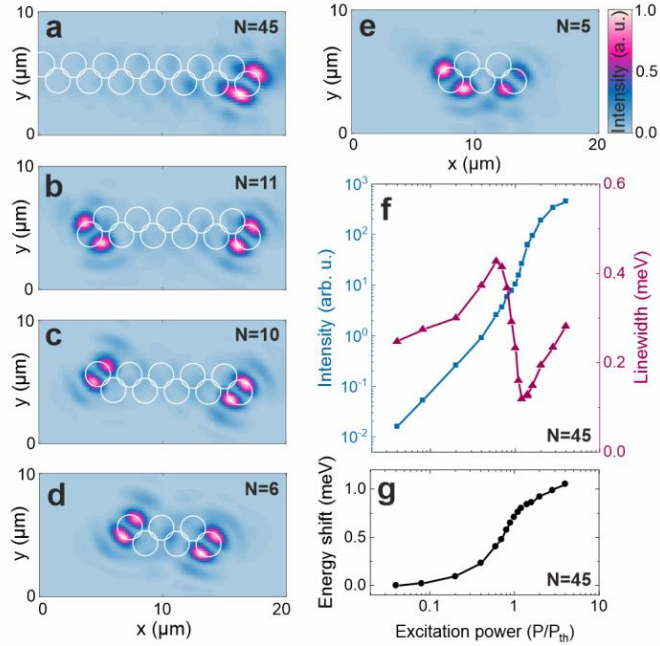

**Figure 2 | Polariton lasing from $P_x$ and $P_y$ topological defect modes in odd and even number SSH-chains.** Real space mode tomographies of SSH defect laser modes in zigzag chains with trap diameters of $d = 2.0\mu m$, a reduced trap distance of $v = 0.9$ and lengths of (**a**) $N = 45$, (**b**) $N = 11$, (**c**) $N = 10$, (**d**) $N = 6$ and (**e**) $N=5$ traps. Inside the topological gap, the emission originates exclusively from these $P_x$ and $P_y$ edge modes ($E_{topo} \sim 1.478 \, meV$). (**f**) Integrated intensity, linewidth and (**g**) mode blueshift of the polariton laser emission from the topological edge state as a function of the excitation power for the right defect of the $N = 45$ chain.

In the following, we systematically investigate the light emission from the polaritonic topological edge defect by exciting a set of chains of different lengths with an elliptical pumping spot of approx. 30 µm by 3 µm, created by a cylindrical lens. We use a non-resonant, pulsed laser operating at a pulse length of 10 ps and a repetition rate of 82 MHz that was tuned to the Bragg minimum at the high-energy side of the stopband. In Fig. 2a, we show the laser-like emission originating from the topological defect of a $N = 45$ zigzag chain comprised of traps with diameters of $d = 2$ μm and a reduced trap distance of $v = a / d = 0.9$. Subsequently, we investigate shortened chains and observe laser emission from defects at both ends of the chain stemming from the same *P* sub-mode for even chain lengths and different *P* sub-modes for odd lengths. For $N = 11$ (b), $N = 10$ (c), $N = 6$ (d), and $N = 5$ (e), we observe topological edge mode lasing. Remarkably, a bulk of mere three sites (e) proves sufficient to keep up the topological nature of the edge defects. The lasing is characterized by a steep increase of output power and a sudden reduction of the linewidth at the lasing threshold (Fig. 2f) as well as a continuous blueshift of the lasing mode (Fig. 2g) stemming from the polariton-polariton interaction.

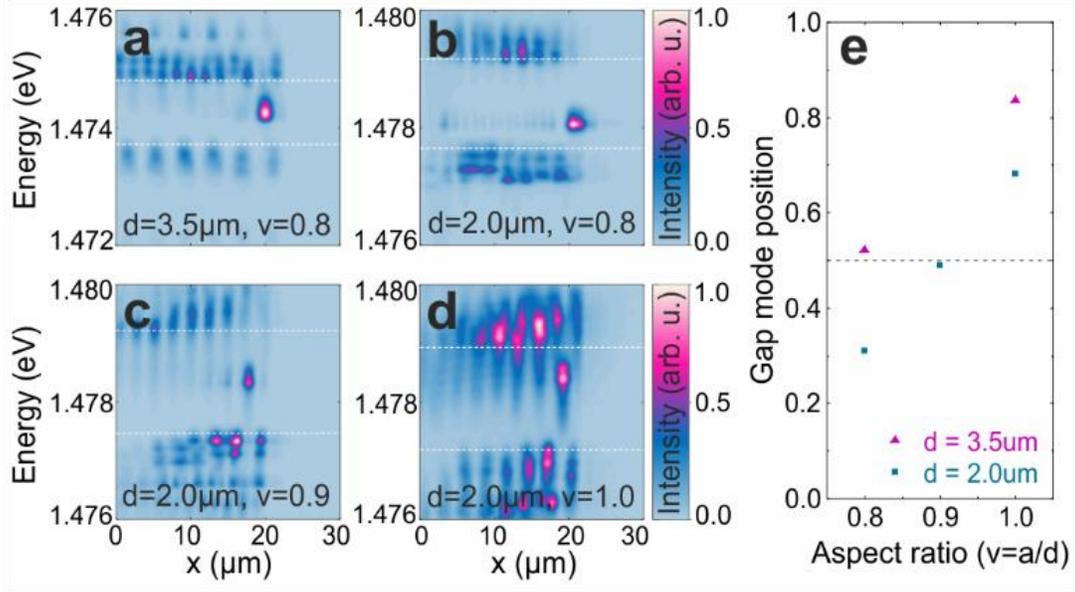

**Figure 3 | Controlling defect mode energy by reduced trap distance *v* and trap diameter *d*.** Real space spectra along a zigzag chain in the linear emission regime showing the topological edge mode positioned in the gap for trap diameters of $d = 3.5\mu m$ and a reduced trap distance of $v = 0.8$ (**a**) as well as trap diameters of $d = 2.0\mu m$ and reduced trap distances of (**b**) $v = 0.8$, (**c**) $v = 0.9$ and (**d**) $v = 1.0$. By varying the trap overlap, the topological mode is shifted with respect to the gap. (**e**) Mode position in the gap with 0 (1) indicating the lower (upper) end of the topological gap as a function of reduced trap distance $v$, evaluated for trap diameters of $d = 2.0$ μm and 3.5 μm.

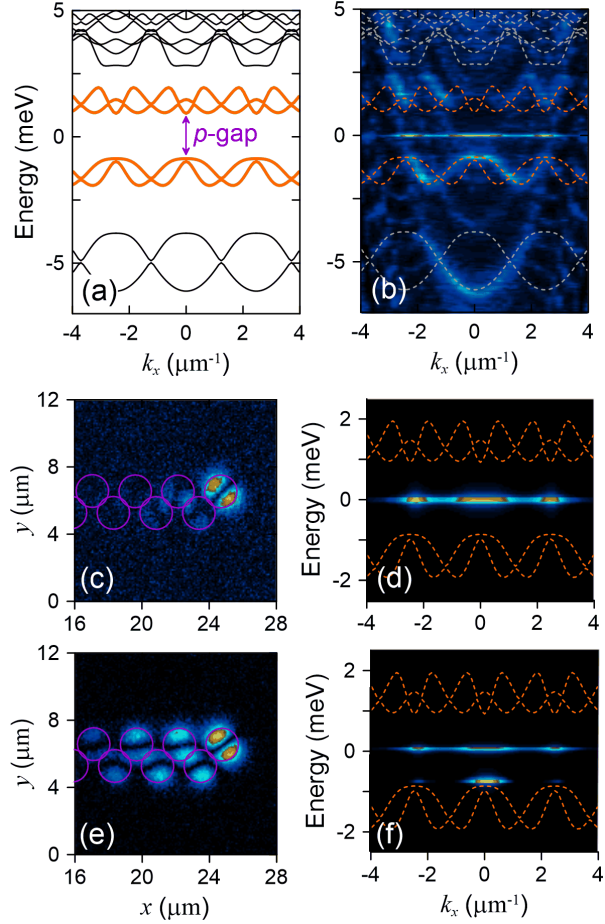

**Figure 4 | Theory of exciton-polariton condensation at the edge of the zigzag chain. a.** Band-gap structure of the infinite chain calculated in the exciton-polariton basis in the single-particle approximation. The forbidden band-gap of interest (*P*-gap) is formed between sub-bands of the *P*-band. Here, the zero energy of the polaritons is chosen in the middle of the *P*-gap. **b.** Spectrum of the truncated zigzag chain calculated below condensation threshold within the modified Gross-Pitaevskii approach. The formation of the edge mode is clearly visible in the *P*-gap of the band-structure. **c.** Spatial intensity profile of the condensate just above condensation threshold for $P_0 = 100 meV\ \mu m^{-2}$. **d.** The spectrum of the condensate shown in (**c**). **e.** The real-space intensity profile of the condensate forming substantially above condensation threshold for $P_0 = 144 meV\ \mu m^{-2}$. **f.** The spectrum of the condensate from (**e**). Among the condensation into the edge mode the condensation into the extended bulk modes of P-band is clearly visible. The chain was pumped incoherently by an optical beam with spatial extension of $16 \mu m \times 6 \mu m$.

Figs. 3a-d display the topological gap and egde mode in real space for a variation of trap diameters and reduced trap distances. Energetically, the topological edge mode is located within a topological gap that well exceeds the mode linewidth by a factor of 7.4. By varying the trap diameter and/or the nearest neighbor distance, we can shift the position of the topological mode within the gap (Fig. 3e) in a continuous and controllable way. This shift is a consequence of deviations from the tight-binding description which we will discuss later on. For two combinations of parameter values, (i) diameter of $d = 2.0$ μm and a reduced trap distance of $v = 0.9$ and (ii) diameter of $d = 3.5$ μm and a reduced trap distance of $v = 0.8$, we find the mode precisely in the center of the gap.

**Generalized Gross-Pitaevskii model describing polariton bandstructure and lasing modes**

In order to comprehensively describe the system as a building block for more elaborated polaritonic topological devices, we aim at a theoretical description of the polariton condensation and consequent laser light emission using a generalized Gross-Pitaevskii (GP) model [36]. The original GP theory has been successfully applied to driven-dissipative systems and typically yields good predictions for polaritons in a planar microcavity in the vicinity of the ground state (see [19] and references therein). However, when studying polariton condensation into non-ground states in deep potential landscapes, this theory has so far shown limited success [37,38]. Here, by using a modified Gross-Pitaevskii approach [36,39,40] as well as calculating the full Bloch modes for the given potential landscape, we find that our theoretical model predicts robust condensation and lasing from the topological defect, in excellent agreement with the experimental findings.

First, we calculate the Bloch modes of the infinitely extended zigzag chain by solving a standard eigenvalue problem (see [11,35,38]) for the coupled excitons and intracavity photons trapped in an external lattice potential. The band-structure of the Bloch states (in the single-particle limit) is depicted in Fig. 4(a). A topologically nontrivial band gap (*P*-gap) forms between two sub-bands within the *P*-band. In the case of a truncated zigzag chain, a topologically protected edge mode is expected to emerge within this gap (see Fig. 4b). To describe our polaritonic zigzag chains, we use a generalized GP equation with realistic sample parameters (see Methods for details). Since the depth of the external potential is comparable with the Rabi splitting, the content of the photonic and excitonic components in polaritons cannot be any longer considered as spatially homogeneous. To take into account this spatial variation, we scale the key system parameters, such as stimulated scattering and polariton-polariton interaction, in accordance with the local fraction of the excitonic component by rescaling them with the Hopfield coefficients. The decay of polaritons from the condensate as well as the reservoir is compensated by an external, off-resonant, time-independent optical pump. The model also accounts for fluctuations of the condensate. In Figure 4b, a radiation spectrum of the truncated zigzag chain under incoherent excitation is displayed. The radiation from the edge mode within the topological gap is clearly visible. Furthermore, at the condensation threshold, the laser emission originates almost exclusively from the topological edge mode (Fig. 4c). An additional indicator is the dispersion of polaritons (Fig. 4 (d)), where we see that, indeed, the topological laser emission is located within the gap in the $P$-band. It is worth mentioning that the pump spot was chosen to be substantially larger than the size of the edge mode and thus extends over several periods of the chain. A further increase of the pumping rate results in the excitation of the extended Bloch modes within the $P$-band of the chain (shown in Fig. 4 e and f). We conclude that the generalized GP model in combination with the full Bloch mode calculation that is based on the actual sample design parameters reproduces our experimental findings remarkably well.

**Temporal coherence measurements and Lindblad master equation modelling**

To further substantiate the topological lasing claim, we have performed measurements of the emission statistics on a $N = 5$ zigzag chain with trap diameters of $d = 3.5\mu m$ and a reduced trap distance of $v = 0.8$ using a Hanbury Brown and Twiss setup with two avalanche photo diodes to measure the second-order coherence function $g^{(2)}(\tau = 0)$ for different excitation powers. In Figures 5a and b, typical correlation measurements at the condensation threshold (corresponding to $P=P_{th}$) and considerably above the threshold are presented. The quantity $g^{(2)}(0)$ is evaluated from the total counts at zero delay, normalized using the side peaks corresponding to counts originating from different laser pulses that are thus uncorrelated. In Fig. 5c, we observe the overshoot (gray area for $P<P_{th}$) that is characteristic for the detector jitter exceeding the coherence time [24,25,42], with the value approaching $g^{(2)}(0)=1$ for higher pumping powers, as expected for a coherent laser. For an excitation power of $P = 3.0$ mW, corresponding to ~4 $P_{th}$, we reach $g^{(2)}(0)$~1.07, which is an excellent result for the coherence of a polariton microlaser. Evidently, there are some limitations on the value of the coherence function due to the polariton-polariton and other scattering processes [22, 43].

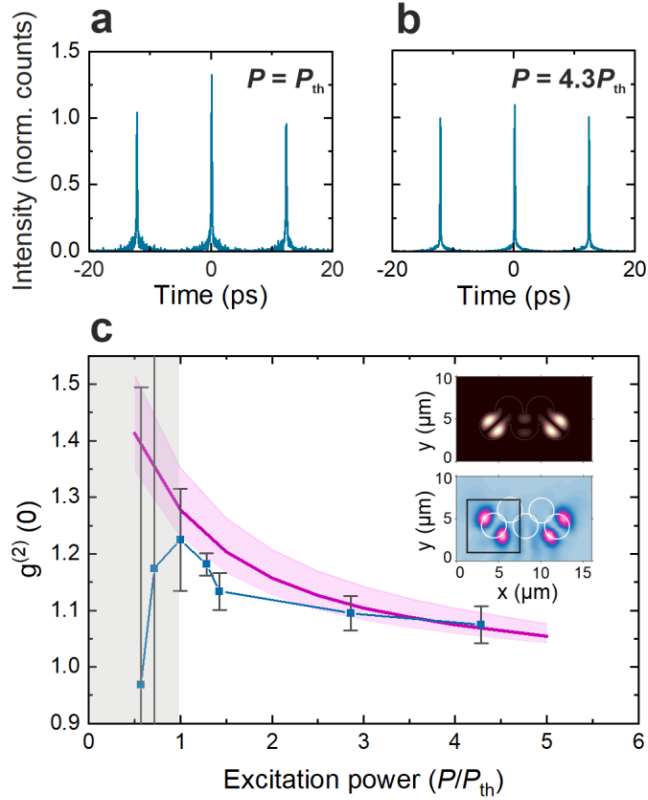

**Figure 5 | Temporal coherence $g^{(2)}(0)$ vs. normalized excitation power: Experiment and theory.**
Exemplary correlation measurements of a topological edge defect of a $N = 5$ zigzag chain at $P = P_{th}$ (**a**) and $P = 4.3P_{th}$ (**b**). The temporal coherence $g^{(2)}$ as a function of excitation power in **c**: experiment (blue) and theory (purple). Inset shows the real-space profile of the modes: theory (upper panel) and experiment (lower panel).
In experiment, the excitation is normalized to a threshold power of $P_{th} \sim 0.7 \ mW$ (see Suppl. for I/O, linewidth and blueshift). The temporal coherence reaches a value of $g^{(2)} \sim 1.07$ for $P \sim 4P_{th}$, indicating a highly coherent polariton lasing state. The theoretical model reproduces the overshoot of $g^{(2)}$ at $P_{th}$ as well as the subsequent decrease of the temporal coherence towards $g^{(2)} \sim 1.00$ in an excellent way. The colored region around the theoretical curve corresponds to a variation of the results of simulation for $\gamma_{k_1,k_2}^{ph} = 0.4 - 0.7 meV$.

To theoretically reproduce the behavior of the temporal coherence function, we use the Lindblad master equation framework [44] to describe dissipative dynamics of the system. The Lindblad equation is one of the general ways to address quantum dissipative evolution and the properties of stationary states. It governs the dynamics of the density matrix $\rho$,

$$\dot{\rho}(t) = -\frac{i}{\hbar}[\hat{\mathcal{H}},\rho(t)] + \sum_s \left(\hat{D}_s \rho(t) \hat{D}_s^\dagger - \frac{1}{2}\{\hat{D}_s^\dagger \hat{D}_s, \rho(t)\}\right),$$

where the first term corresponds to Hamiltonian dynamics and the second term is a sum over all the dissipative channels $s$ with $\hat{D}_s$ the dissipative operators, also called *jump operators* (for details, see Methods). We model the coherent dynamics with a two-component Hamiltonian [45], consisting of a single-particle component $\hat{\mathcal{H}}_p$ and the multi-polariton interaction $\hat{\mathcal{H}}_i$. This interaction arises from photon-exciton hybridization. Using the finite difference method, we diagonalize the single-polariton Hamiltonian of the system shown in Fig. 1, and we find the eigenvalues consisting of the energy and the inverse time of the modes $\{E_k - i\gamma_k\}$ for each eigenstate, denoted by $k$. The non-Hermitian components of the eigenvalues are included in the dissipation operators' rates and we additionally include polariton-polariton interactions phenomenologically. For the incoherent component of the evolution, we utilize a model developed for the treatment of polaritonic nonequilibrium systems in [46]. This model has previously been successful in describing the coherence properties of single micropillar lasers with diameters down to 6 μm [25]. It describes a generic bosonic system under an incoherent pumping and in contact with a thermal reservoir which causes dissipative phonon-mediated scattering of polaritons. With this setup, we can solve the master equation exactly by diagonalization of the Lindbladian matrix, and calculate the second-order temporal coherence function for the *k*th mode:

$$g_k^{(2)}(\tau) = \frac{\left\langle a_k^\dagger(t) a_k^\dagger(t+\tau) a_k(t+\tau) a_k(t) \right\rangle}{\left\langle a_k^\dagger(t) a_k(t) \right\rangle^2},$$

where $k$ corresponds to the *P*-mode, and $\tau$ is the time delay. The averaging is done over the stationary state, i.e. at $\rho(t \to \infty)$. At zero time delay it corresponds to the coherence function measured using the Hanbury Brown and Twiss setup. The results are presented in Fig. 5c. The theory nicely reproduces the experimental data in the region of not too small pumping intensities, i.e. for $P \geq P_{th}$. The error bars are calculated from the fluctuation of the $g_k^{(2)}(\tau)$ side peaks for finite delay.

**Magnetic field measurement to substantiate the hybrid light-matter nature**

After we have unequivocally demonstrated the coherent, laser-like light emission from the topological edge mode, we continue by substantiating that this light is indeed emitted in the strong-coupling regime, thus reflecting the hybrid light-matter nature of the system. For that, we apply a magnetic field to the sample in Faraday geometry and measure the emission spectra. Here, the field is varied from 0T to 5T in steps of 1T and we measure the Zeeman splitting and the threshold. As plotted in Fig. 6a, a Zeeman splitting of approximately 3.9 μeV/T, corresponding to 19.6 μeV at 5T, was found for the topological mode. Under the reasonable assumption that the observed splitting is a product of the exciton Zeeman splitting of 355 μeV (measured on uncoupled quantum wells), and the excitonic fraction of the polariton (described by the Hopfield coefficient of $|X|^2 = 0.06$ for the topological edge mode), a splitting of 22.3 µeV is expected at 5T. This estimation agrees well with the observed Zeeman splitting. In addition, we observe a reduction of the condensation threshold when increasing the magnetic field. In Figure 6b, the characteristic decrease in linewidth at the threshold, which serves to visualize the reduction of the condensation threshold, is depicted (see also the corresponding evaluations of the PL intensity and the energy shift in the Supplemental Material). Such a behavior is common for polaritons and can be attributed to an interplay of a favorable change of detuning, spin polarization enhancing exciton-exciton scattering, and enhancement of the exciton oscillator strength with the increase of the magnetic field [47,48].

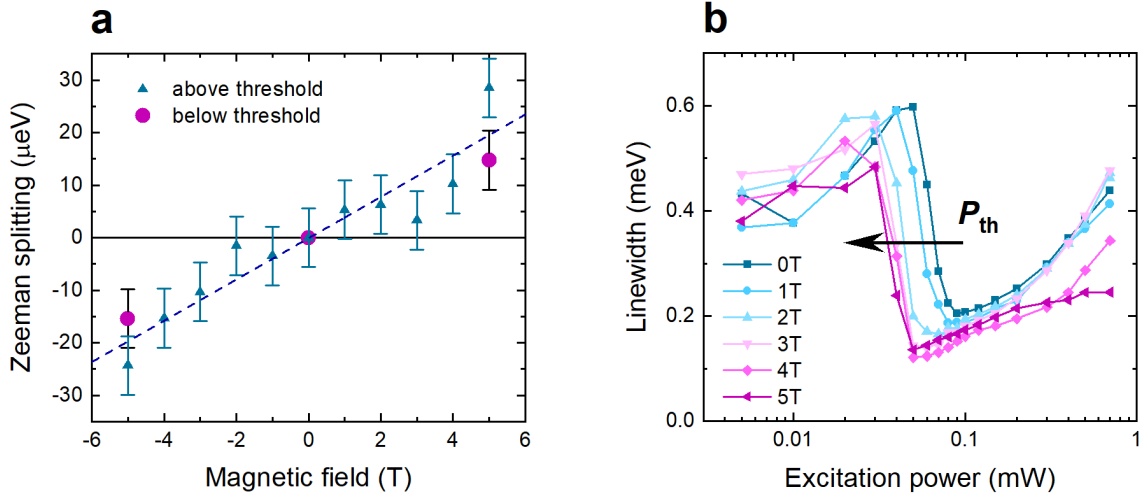

**Figure 6 | Zeeman splitting and lasing threshold reduction of a topological polariton laser mode in a magnetic field. a,** Zeeman splitting of the topological polariton edge mode as a function of applied magnetic field in Faraday geometry. A linear increase in Zeeman-splitting due to the excitonic matter part of the polaritons is observed. **b,** Threshold behavior of a topological polariton edge mode visualized by the characteristic decrease in linewidth as a function of the excitation power with varying magnetic field strength. The threshold of the topological polariton laser decreases with increasing magnetic field.

**Discussion**

In conclusion, we have demonstrated polariton lasing from a topological defect in an SSH chain. We have performed measurements of the emission from the system and investigated the properties of the emitted light. We have also used an external magnetic field to confirm the strong light-matter coupling regime and thus the hybrid nature of lasing by measuring Zeeman splitting. To support the experimental findings, we have used a comprehensive theoretical model that is able to correctly predict and describe the condensation in the topological defect by means of a generalized GP equation. Moreover, we have studied the coherence properties of the topological laser using the Lindblad master equation framework. The relative simplicity of the SSH model serves as a benefit that enables an extensive description while still allowing us to use these findings to implement more complex, two-dimensional topological lasing schemes. In such schemes, the necessity for a topological laser mode to propagate in combination with its topological protection [14] ensures single mode, coherent lasing that is unaffected by local disorder. In this context, our work is an important step towards potential applications of topological concepts in the realms of semiconductor laser physics, in particular as electrical injection of polaritonic devices has already been demonstrated [49,50].

**Acknowledgements**

The Würzburg group acknowledges support from the DFG through the Würzburg-Dresden Cluster of Excellence on Complexity and Topology in Quantum Matter "*ct.qmat*" (EXC 2147, project-id 39085490) and the doctoral training program "Elitenetzwerk Bayern". S.H. acknowledges support by the EPSRC *"Hybrid Polaritonics"* Grant (EP/M025330/1). T.H. gratefully acknowledges support by the German Academic Scholarship Foundation. M.S., I.V. and I.G.S acknowledge the support by the Institute for Basic Science in Korea (Project No.~IBS-R024-D1).

**Authors contributions**

S.K., C. S. and S.H. initiated the study and guided the work. T.H.H., J.B., P.G., M.E., C.S. and S.K. designed and fabricated the device. T.H.H. performed optical measurements. T.H.H. and S.K. analyzed and interpreted the experimental data. M.S., I.V., O.A.E., U.P., and I.G.S. developed the theory. S.K., T.H.H., O.A.E and I.G.S. wrote the manuscript, with input from all coauthors.

**Additional information**

**Competing financial interests:** The authors declare no competing financial interests.

**Correspondence and requests** for materials should be addressed to Sebastian Klembt (sebastian.klembt@physik.uni-wuerzburg.de) or Sven Höfling (sven.hoefling@physik.uni-wuerzburg.de).

**Methods**

In our sample, the zigzag chains are implemented using the etch and overgrowth (EnO) approach. Here, the growth is interrupted after the bottom distributed Bragg reflector (DBR) and the cavity are finished and the traps are defined by electron beam lithography and a subsequent etching step. The zigzag chains were etched into the GaAs λ-cavity to a depth of approximately 10 nm, resulting in a confinement potential of 11.5 meV. The cavity is embedded between two $Al_{0.1}Ga_{0.9}As$/AlAs DBRs with 27 and 33.5 mirror pairs in the top and bottom DBRs, respectively, resulting in a quality factor of $Q = 7200$. There is a strong light-matter coupling between (i) the excitonic resonances of three 16 nm thick $In_{0.04}Ga_{0.96}As$ quantum wells located at the antinode of the electric field of the cavity mode and (ii) the cavity resonance. It manifests itself in a pronounced anti-crossing behavior of the eigenmodes (see Supplemental Material) with a Rabi splitting of 4.5 meV. For the experiments presented in this work, zigzag chains featuring a moderately negative photon-exciton detuning of approximately -12 meV were selected.

For the Bloch mode calculation, we use the generalized GP equation

$$i\hbar \frac{\partial \Psi(\mathbf{r},t)}{\partial t} = \left[ -\frac{\hbar^2}{2m_{\text{eff}}}\nabla^2 - \frac{i\hbar\gamma_C}{2} + V_{\text{ext}}(\mathbf{r}) + g_C(\mathbf{r})|\Psi(\mathbf{r},t)|^2 \right.$$
$$\left. + \left(g_r(\mathbf{r}) + \frac{i\hbar R(\mathbf{r})}{2}\right)n_R(\mathbf{r},t) \right]\Psi(\mathbf{r},t) + i\hbar\frac{d\Psi_{\text{st}}(\mathbf{r},t)}{dt}$$

$$\frac{\partial n_R(\mathbf{r},t)}{\partial t} = -(\gamma_R + R(\mathbf{r})|\Psi(\mathbf{r},t)|^2)n_R(\mathbf{r},t) + P(\mathbf{r})$$

where $\Psi(\mathbf{r},t)$ is a collective wave function of lower-branch polaritons in the plane of the microcavity with $\mathbf{r} \equiv \{x,y\}$, and $n_R(\mathbf{r},t)$ is a high-energy exciton reservoir density. We use typical parameters: Rabi splitting $\hbar\Omega = 5.51 meV$, photon-exciton detuning $\hbar(\omega_{\text{cavity}} - \omega_{\text{exciton}}) \equiv \hbar\Delta_0 = -13.5 meV$, and the effective mass of intracavity photons $m_C = 36 \cdot 10^{-6} m_e$ with the free electron mass $m_e$. Then, the effective mass of lower-branch polaritons can be estimated by $m_{\text{eff}} = 2m_C\sqrt{\Delta_0^2 + \Omega^2}/(\sqrt{\Delta_0^2 + \Omega^2} - \Delta_0)$. The effective trapping potential for polaritons is given by the expression $V_{\text{ext}}(\mathbf{r}) = \frac{1}{2}(V(\mathbf{r}) + \sqrt{(\Delta_0 - V(\mathbf{r}))^2 + \Omega^2} - \sqrt{\Delta_0^2 + \Omega^2}$, where the external potential for the intracavity photons is determined by the zigzag chain consisting of the mesas in the form of super Gauss $V(\mathbf{r}) = \sum_i V_0 \exp\left(-\frac{|\mathbf{r}-\mathbf{r}_i|^{50}}{d^{50}}\right)$ with the potential depth $V_0 = 11.5 meV$ and mesa diameters $d = 2\mu m$ (centered at $\mathbf{r}_i$).

Since the depth of the external potential is comparable with Rabi splitting, the content of the photonic and excitonic components in polaritons cannot be any longer considered as spatially homogeneous. To take into account the spatial variation, we scale the key system parameters, such as stimulated scattering $R(\mathbf{r})$ and polariton-polariton interaction $g_C(\mathbf{r})$ and $g_R(\mathbf{r})$, in accordance with the local fraction of the excitonic component by rescaling them with the Hopfield coefficients:

$$|C(\mathbf{r})|^2 = \frac{1}{2}\left(1 - \frac{\Delta_0 - V(\mathbf{r})}{\sqrt{(\Delta_0 - V(\mathbf{r}))^2 + \Omega^2}}\right), \quad |X(\mathbf{r})|^2 = 1 - |C(\mathbf{r})|^2.$$

Then, the stimulated scattering from the reservoir to (coherent) polaritons becomes $R(\mathbf{r}) = R_0|X(\mathbf{r})|^2$ with the fitting parameter $\hbar R_0 = 0.01 meV\ \mu m^2$. The strengths of polariton-polariton and polariton-reservoir interactions are characterized by the expressions $g_C(\mathbf{r}) = g|X(\mathbf{r})|^4$ and

$g_R(r) = g|X(r)|^2$, respectively, where $\hbar g = 0.01 meV\ \mu m^2$ describes the effective interaction between bare excitons. The constants $\hbar\gamma_C = 0.1 meV$ and $\hbar\gamma_R = 0.2 meV$ stand for the decay rates of condensed polaritons and reservoir, respectively. These losses are compensated by an external off-resonant time-independent optical pump with the injection rate $P(r)$.

The model also accounts for fluctuations of the condensate. The corresponding terms are derived within the truncated Wigner approximation [40,41], giving $d\Psi_{st}(r_l) = \sqrt{(\gamma_C + R(r_l)n_R(r_l))/(4\delta x \delta y)}\, dW_l$. Here, $dW_l$ is a Gaussian random variable characterized by the correlation functions $\langle W_l^* dW_j\rangle = 2\delta_{l,j} dt$ and $\langle dW_l dW_j\rangle = 0$ where $l, j$ are discretization indices.

The temporal coherence function was theoretically reproduced using the Lindblad equation which governs the dynamics of the density matrix $\rho$,

$$\dot{\rho}(t) = -\frac{i}{\hbar}[\hat{\mathcal{H}}, \rho(t)] + \sum_s \left(\hat{D}_s \rho(t) \hat{D}_s^\dagger - \frac{1}{2}\{\hat{D}_s^\dagger \hat{D}_s, \rho(t)\}\right),$$

where the first term corresponds to Hamiltonian dynamics and the second term is a sum over all the dissipative channels $s$ with $\hat{D}_s$ the dissipative operators, also called *jump operators*. We model the coherent dynamics with a two-component Hamiltonian, consisting of a single-particle component $\hat{\mathcal{H}}_p$ and the multi-polariton interaction $\hat{\mathcal{H}}_i$. In real space, the single-particle Hamiltonian it reads

$$\hat{\mathcal{H}}_p \begin{bmatrix} \psi \\ \phi \end{bmatrix} = \begin{bmatrix} -\frac{\hbar^2}{2m_C}\nabla^2 + V(x,y) - \frac{i\hbar}{2\tau_{cp}} & \frac{\hbar\Omega}{2} \\ \frac{\hbar\Omega}{2} & -\frac{\hbar^2}{2m_X}\nabla^2 - \hbar\Delta_0 \end{bmatrix} \begin{bmatrix} \psi \\ \phi \end{bmatrix},$$

where $\psi$ and $\phi$ are the wave functions of the cavity photons and excitons, respectively. The corresponding effective masses are $m_C = 2.5 \times 10^{-5} m_e$ and $m_X = 10^5 m_C$. The Rabi splitting $\Omega$ and detuning $\Delta_0$ between the cavity photon and the exciton dispersions are taken to be the same as in the experiment. The potential of the barrier $V(x, y)$ is shaped as the SSH chain with an amplitude of $V = 2000 meV$; $\tau_{cp} = 7.3 ps$ is the lifetime of the cavity photons. Using the finite difference method, we diagonalize the single-polariton Hamiltonian of the system shown in Fig. 1, and we find the eigenvalues consisting of the energy and the inverse time of the modes $\{E_k - i\gamma_k\}$ for each eigenstate (each denoted by $k$, with energies sorted in the ascending order). For the analysis of $g^{(2)}$ in the P-edge state, we include three different states in the system. We consider the states k=10, 58 and 59 which correspond to the lower P-edge state and two states just below the bottleneck region. Based on this, we transform the problem into a discrete bosonic model with the states defined by the eigenstates of $\widehat{\mathcal{H}}_p$. Thus, in the new basis, the Hermitian part of the Hamiltonian reads $\widehat{\mathcal{H}}_p = \sum_k E_k \hat{a}_k^\dagger \hat{a}_k$, where $\hat{a}_k^\dagger$ is $k$th mode creation operator. The maximum number of excitations is $N_{max} = 20$ and $N_{max} = 5$ for the edge state and the other states, respectively. The non-Hermitian components of the eigenvalues are later used in the dissipation operators' rates. Additionally, we include polariton-polariton interaction phenomenologically as $\widehat{\mathcal{H}}_i = U \sum_k \hat{a}_k^\dagger \hat{a}_k^\dagger \hat{a}_k \hat{a}_k$, where $U$ is a particle interaction strength. We calculate its value using the condition $\frac{U}{N_{max}} = 1 meV$ and disregarding any coupling between different polariton modes. Hence, the coherent evolution is governed by the total model Hamiltonian

$$\widehat{\mathcal{H}} = \widehat{\mathcal{H}}_p + \widehat{\mathcal{H}}_i = \sum_k E_k \hat{a}_k^\dagger \hat{a}_k + U \sum_k \hat{a}_k^\dagger \hat{a}_k^\dagger \hat{a}_k \hat{a}_k.$$

For the incoherent component of the evolution, we utilize a model developed for the treatment of polaritonic nonequilibrium systems that describes a generic bosonic system under an incoherent pumping and in contact with a thermal reservoir which causes dissipative phonon-mediated scattering of polaritons. The jump operators describing pump and losses of the system read

$$\widehat{D}_k^{(\text{pump})} = \hat{\mathcal{J}}_k^+ = \sqrt{\gamma_k \eta_\text{p}(E_k)}\hat{a}_k^\dagger,$$

$$\widehat{D}_k^{(\text{loss})} = \hat{\mathcal{J}}_k^- = \sqrt{\gamma_k [\eta_\text{p}(E_k) + 1]}\hat{a}_k,$$

where $\eta_\text{p}(E_k) = \left(\exp(E_k/k_B T_\text{p}) - 1\right)^{-1}$ is the Bose distribution function of the pumping reservoir controlled by its temperature $T_\text{p}$. We consider all the modes as being lossy, thus $\widehat{D}_k^{(\text{loss})}$ implies summation over all the values of $k$. As for the pumping, it is only applied to the highest-energy mode. We define the pumping as $P = \gamma_k \eta_k$ where $\gamma_k$ is the inverse lifetime of the maximum-energy state, and the pumping threshold is defined by the lifetime of the $P$-edge state which is $P_\text{th} = \gamma_{k_P}$, where $k_P$ denotes the $P$-state index. Similarly, the phonon-mediated relaxation is governed by the following jump operators

$$\widehat{D}_k^{(\text{exc})} = \hat{\mathcal{J}}_{k_1,k_2}^+ = \sqrt{\gamma_{k_1,k_2}^{\text{ph}} \eta_{\text{ph}}(E_{k_1} - E_{k_2})}\hat{a}_{k_1}^\dagger a_{k_2},$$

$$\widehat{D}_k^{(\text{dump})} = \hat{\mathcal{J}}_{k_1,k_2}^- = \sqrt{\gamma_{k_1,k_2}^{\text{ph}} [\eta_{\text{ph}}(E_{k_1} - E_{k_2}) + 1]}\hat{a}_{k_1} a_{k_2}^\dagger,$$

where $\gamma_{k_1,k_2}^{\text{ph}}$ is the scattering rate, and the $\eta_{\text{ph}}(E)$ is Bose distribution function of the phonon reservoir, characterized by the temperature $T_\text{ph} = 5$ K, which is effectively the system temperature. These processes imply summation only over the pairs of states with $E_{k_1} > E_{k_2}$, so the first set of jump operators corresponds to incoherent excitations, and the second set – to incoherent damping. The interaction strength between the phonons and polaritons is set to $\hbar \gamma_{k_1,k_2}^{\text{ph}} = 0.55 meV$.

With this setup, we can solve the master equation exactly by diagonalization of the Lindbladian matrix, and calculate the second-order temporal coherence function for the $k$th mode:

$$g_k^{(2)}(\tau) = \frac{\langle a_k^\dagger(t) a_k^\dagger(t+\tau) a_k(t+\tau) a_k(t) \rangle}{\langle a_k^\dagger(t) a_k(t) \rangle^2},$$

where $k$ corresponds to the $P$-mode, and $\tau$ is the time delay. The averaging is done over the stationary state, i.e. at $\rho(t \to \infty)$. At zero time delay it corresponds to the coherence function measured using the Hanbury Brown and Twiss setup.